\documentclass[9pt]{article}
\usepackage{spconf}

\usepackage{amsmath,amsfonts,graphicx}
\usepackage{amssymb,textcomp,mathtools}
\usepackage{bm,upgreek,algorithm,hyperref}
\usepackage{multirow,booktabs,hhline,array}
\usepackage{cite,url,makecell,setspace}
\usepackage{xcolor}
\pdfoutput=1

\def\thline{\noalign{\hrule height 1.0pt}}

\renewcommand{\vec}[1]{\bm{\mathrm{#1}}}

\title{On the Design and Training Strategies for RNN-based Online Neural Speech Separation Systems}
\name{Kai~Li$^{\dagger}$$^{\ddagger}$\sthanks{Work done during internship at Tencent AI Lab.}, Yi~Luo$^\ddagger$}
\address{$^\dagger$Department of Computer Science and Technology, BNRist, Tsinghua University, China \\ $^\ddagger$Tencent AI Lab, Shenzhen, China \\
lk21@mails.tsinghua.edu.cn, oulyluo@tencent.com}

\begin{document}
\maketitle
\ninept

\begin{abstract}
While the performance of offline neural speech separation systems has been greatly advanced by the recent development of novel neural network architectures, there is typically an inevitable performance gap between the systems and their online variants. In this paper, we investigate how RNN-based offline neural speech separation systems can be changed into their online counterparts while mitigating the performance degradation. We \textit{decompose} or \textit{reorganize} the forward and backward RNN layers in a bidirectional RNN layer to form an \textit{online path} and an \textit{offline path}, which enables the model to perform both online and offline processing with a same set of model parameters. We further introduce two training strategies for improving the online model via either a pretrained offline model or a multitask training objective. Experiment results show that compared to the online models that are trained from scratch, the proposed layer decomposition and reorganization schemes and training strategies can effectively mitigate the performance gap between two RNN-based offline separation models and their online variants.
\end{abstract}

\section{Introduction}
\label{sec:introduction}
Recent advances in neural speech separation systems mostly assume an offline processing pipeline. This means that either the separation requires multiple stages where the global information needs to be gathered in at least one of the stages \cite{isik2016single, zeghidour2020wavesplit}, or the network architecture takes the entire utterance as input \cite{nachmani2020voice, subakan2020attention, afrcnn}. On the other hand, streaming speech separation systems are important in applications such as daily communications, telecommunication systems, or streaming automatic speech recognition (ASR) systems. A straightforward way to modify an offline system to an online system is to change the network modules that make use of the global information, e.g. bidirectional recurrent neural network (Bi-RNN) layers, to modules that only make use of the history information, e.g. unidirectional RNN (Uni-RNN) layers \cite{luo2018tasnet, tan2018convolutional, han2019online}. However, there is typically a significant performance gap between an offline system and its online variant \cite{luo2019conv, liu2020causal}.

Existing methods for improving the performance of online neural speech separation systems include the application of \textit{look-ahead window} and \textit{training objectives}. Compared to strictly online neural speech separation systems, adding a look-ahead window typically greatly improves the separation performance at the cost of a higher minimal system latency \cite{wilson2018exploring, sonning2020performance}. Moreover, another advantage of using a look-ahead window is that there is typically no need to change the detailed model architectures. Training objectives for online neural speech separation systems include teacher-student training \cite{hinton2015distilling}, where the output from a pretrained offline model can be used to guide the training of the online model \cite{aihara2019teacher}, and predictive training, where the online neural network not only performs separation but also predicts the next few frames which can further be used as auxiliary input for the separation of next frames \cite{li2019listening}.

In this paper, we propose another method for improving online neural speech separation systems by investigating novel model design and training strategies. Specifically, we focus on RNN-based systems where the switch from offline systems to online systems can be done by modifying Bi-RNN layers to Uni-RNN layers. We propose to \textit{decompose} or \textit{reorganize} a Bi-RNN layer into an \textit{online path} and an \textit{offline path}. In the decomposition scheme, the backward RNN is ignored in the online path and only the output from the forward LSTM is used. In the reorganization scheme, the backward RNN takes the time-reversed sequence as input in the offline path, and takes the original sequence as input in the online path. These two schemes allows a Bi-RNN layer to perform both online and offline processing with a same set of model parameters. We further introduce two training strategies via either initializing the online system with a pretrained offline system or training an offline system with both online and offline outputs. Experiment results show that the proposed layer decomposition and reorganization schemes, together with the training strategies, improve the performance of online neural speech separation systems by up to 1 dB in terms of scale-invariant signal-to-distortion ratio (SI-SDR).


\begin{figure*}[!ht]
	\small
	\centering
	\includegraphics[width=2\columnwidth]{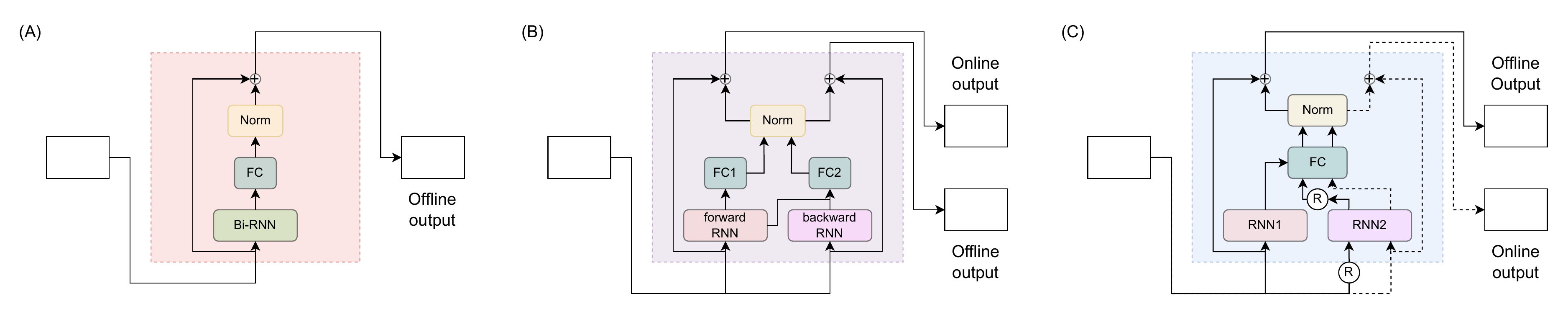}
	\caption{(A) A standard Bi-RNN layer with normalization operation and residual connection. (B) Decomposition of a Bi-RNN layer into an online path and an offline path. The backward LSTM is ignored in the online path. (C) Reorganization of a Bi-RNN layer into an online path and an offline path. One LSTM takes reversed sequence as input in the offline path, and takes original sequence as input in the online path.}
	\label{fig:pipeline}
\end{figure*}

\section{Bi-RNN Layer Decomposition and Reorganization for Online Speech Separation}
\label{sec:model}
\subsection{Online and Offline Decomposition and Reorganization of Bi-RNN Layers}

Figure~\ref{fig:pipeline} (A) shows a conventional design of a Bi-RNN layer with normalization and residual connection. A sequential feature is sent to the Bi-RNN layer to generate another sequential feature, and a fully-connected (FC) layer is applied to each of the time step for reshaping the dimension of the transformed feature back to the dimension of the input feature. A normalization operation is then applied to the output of the FC layer, and a residual connection is added between the input to the Bi-RNN layer and the normalized output.

Figure~\ref{fig:pipeline} (B) shows how the Bi-RNN layer can be \textit{decomposed} into an \textit{online path} and an \textit{offline path}. We consider the forward RNN layer and the backward RNN layer in the bidirectional RNN layer independently, and \textbf{use the forward RNN layer to generate another output}. The process done by the forward RNN layer is referred to as the \textit{online path}, where another FC layer is applied to the output of the forward RNN layer to match the feature dimension to the input of the forward and backward RNN layers. The standard process done by the Bi-RNN layer is referred to as the \textit{offline path}. The normalization operation is shared by the outputs of both the forward RNN layer and the original Bi-RNN layer (followed by their corresponding FC layers, respectively), and the residual connections are added between the input to the Bi-RNN layer to the outputs from both the online path and the offline path. For deep RNN architectures, the online and offline paths can be treated independently across layers, which can be implemented by creating two copies of the input sequential feature to the first Bi-RNN layer and select online or offline paths throughout the entire network.

Figure~\ref{fig:pipeline} (C) shows how the Bi-RNN layer can be \textit{reorganized} into an online path and an offline path. The \textit{RNN1} is identical to the forward RNN layer in Figure~\ref{fig:pipeline} (B) that performs standard sequence modeling. The \textit{RNN2} now takes two different sequences as input, one is the time-reversed sequence and another one is the original sequence. When the time-reversed sequence is used and the output of RNN2 is also time-reversed, it is identical to the backward RNN layer in Figure~\ref{fig:pipeline} (B). When the original sequence is used, the Bi-RNN layer consists of two forward RNN layers and thus becomes unidirectional. Since now the outputs from the two RNN layers are used in both online and offline paths, the FC layer for dimension matching can be shared by both paths. Similar to Figure~\ref{fig:pipeline} (B), a normalization operation shared by a residual connection is added to both the online and offline paths to generate two outputs.

By such layer decomposition and reorganization schemes, a same model can perform both online and offline processing depending on which path the model selects, which allows us to select from two possible strategies for model training:
\begin{enumerate}
    \item We first pretrain an offline model with Bi-RNN layers, and use the layers in the Bi-RNN layers for initialization in its online counterpart. For the decomposition scheme, only the forward RNN layer is used; for the reorganization scheme, both RNN layers are used.
    \item We train an offline model with Bi-RNN layers with training objectives applied to both the outputs from the online path and the offline path.
\end{enumerate}
The first one can be treated as a \textit{finetuning} method or a \textit{curriculum learning} process \cite{bengio2009curriculum}, and the second one can be treated as a \textit{multitask learning} method.

Although any RNN-based network architectures can apply such decomposition or reorganization scheme, here we take dual-path RNN (DPRNN) block \cite{luo2020dual}, an RNN-based sequential modeling framework for time-domain speech separation, as an example and describe how it can be modified to its online variant. For a quick recap of a standard DPRNN block, it splits a two-dimensional sequential feature into a sequence of overlapped chunks, and the chunks are concatenated to form a three-dimensional tensor. An \textit{intra-chunk RNN} module is first applied to all the chunks to model the intra-chunk dependencies of the entire sequence, and then an \textit{inter-chunk RNN} module is applied across the chunks to model the global dependencies of the entire sequence. For offline processing, both intra-chunk and inter-chunk RNNs are Bi-RNNs. Here we keep the intra-chunk Bi-RNN unchanged and modify the inter-chunk Bi-RNN via the same way as Figure~\ref{fig:pipeline} (B) or (C) to turn the offline DPRNN block into a chunk-online DPRNN block whose minimum latency is decided by the chunk length. For stacked DPRNN blocks, the online path is defined by the intra-chunk Bi-RNN modules and the forward RNN layer in the inter-chunk Bi-RNN modules.

\subsection{Normalization Operation}

Deep RNN architectures typically contain normalization operations. Global normalization operations, such as global layer normalization (gLN) \cite{luo2019conv}, is used in several RNN-based neural speech separation systems \cite{luo2018tasnet, luo2020dual}. For online processing, the normalization operation needs to be modified to satisfy the latency constraint of the entire system. Here we adopt the cumulative layer normalization (cLN) \cite{luo2019conv} to calculate the running statistics:
\begin{align}
cLN(\vec{f}_k) &= \frac{\vec{f}_k - E[\vec{f}_{t\leq k}]}{\sqrt{Var[\vec{f}_{t\leq k}] + \epsilon}} \odot \gamma + \beta \\
E[\vec{f}_{t\leq k}] &= \frac{1}{Nk} \sum_{Nk}{\vec{f}_{t\leq k}} \\
Var[\vec{f}_{t\leq k}] &= \frac{1}{Nk} \sum_{Nk}{(\vec{f}_{t\leq k} - E[\vec{f}_{t\leq k}])^2}
\label{eqn:cLN}
\end{align}
where $\vec{f}_k \in \mathbb{R}^{N\times 1}$ is the $k$-th frame of the entire feature $\vec{F}$, $\gamma, \beta \in \mathbb{R}^{N\times 1}$ are trainable parameters, and $\vec{f}_{t\leq k} \in \mathbb{R}^{N\times k}$ corresponds to the feature of $k$ frames $[\vec{f}_1, \ldots, \vec{f}_k]$. For stacked Bi-RNN layers, cLN can directly replace the gLN operations in all the layers, while for stacked DPRNN blocks, we still use gLN in the intra-chunk RNN modules and use cLN in the inter-chunk RNN modules. Given that the normalization operation is typically also applied to the output of a signal encoding module, e.g. magnitude spectrogram generated by short-time Fourier transform (STFT) or a learnable representation generated by a 1-D convolutional layer, we also need to replace gLN by cLN for normalizing the signal encoder output.

\subsection{Related Works}

The model configuration of using a same model for both online and offline processing has been investigated in the task of streaming ASR \cite{yu2020dual}. The proposed \textit{Dual-mode ASR} system uses a same model to perform both streaming and full-context ASR via adjusting certain operations such as the convolution operation, average pooling operation, and the self-attention calculation. More specifically, ``noncausal'' convolution operations that takes future information as input can be modified to a ``causal'' convolution operation by zero-padding at the past, ``noncausal'' average pooling operation can be modified to a ``causal'' average pooling operation by only considering the context window in the past, and ``noncausal'' self-attention calculation can be modified to a ``causal'' self-attention calculation by adjusting the masking matrix for attention, e.g. to perform time-restricted self-attention \cite{vaswani2017attention, povey2018time}. However, the prior works on modifying RNN-based architectures to perform such dual-mode processing is limited.

The idea of using different sub-modules in a neural network to perform different tasks can be connected to the prior works on dynamic neural networks \cite{han2021dynamic} and slimmable neural networks \cite{yu2018slimmable}. The layer decomposition scheme can be connected to the \textit{path selection} mechanism in dynamic neural networks \cite{liu2018dynamic}, where one out of multiple sub-modules is dynamically selected from the entire network to process the current input. Unlike dynamic neural networks that requires an additional path-selection module, our layer decomposition scheme is a simplified configuration where the path selection in online and offline paths are deterministic. On the other hand, slimmable neural networks use different width of a same layer to perform model training, so that the actual model width can be selected by the hardware constraints and requirements during deployment. The layer decomposition scheme can thus be connected to slimmable neural networks in the way that the forward and backward RNN layers can be viewed as the decomposition of a larger RNN meta-layer with 50\% width in each layer. However, slimmable neural networks were typically investigated in convolutional or fully-connected architectures, while their application in RNN-based architectures has not been well studied yet.

The \textit{RNN2} layer in the layer reorganization scheme takes time-reversed sequence as input in offline path and takes original sequence as input in online path. The idea of using a same model to take both original and time-reversed sequence as input has been investigated in a prior work on the speech enhancement model for noise-robust ASR \cite{chao2021tenet}, where a noise waveform is time-reversed as a data augmentation scheme for improving the denoising performance. However, the system only considered the reversal of the input waveform as a training method and did not consider joint modeling of online and offline paths with a same model.

\section{Experiment configurations}
\label{sec:config}
\subsection{Dataset}

We evaluate the proposed Bi-RNN decomposition method and training strategies on a simulated noisy reverberant two-speaker dataset. 20000, 5000 and 3000 4-second long utterances are simulated at 16k Hz for training, validation and test sets, respectively. For each utterance, two speech signals and one noise signal are randomly selected from the 100-hour Librispeech subset \cite{panayotov2015librispeech} and the 100 Non-speech Corpus \cite{web100nonspeech}, respectively. The overlap ratio between the two speakers is uniformly sampled between 0\% and 100\%, and the two speech signals are rescaled to a random relative signal-to-noise-ratio (SNR) between 0 and 5 dB. The relative SNR between the speech mixture and the noise is randomly sampled between 10 and 20 dB. The source signals are then convolved with the room impulse response filters simulated by the image method \cite{allen1979image} using the gpuRIR toolbox \cite{diaz2020gpurir}. The clean reverberant source signals are used as the training target for all experiments.

\subsection{Model configurations}

We build networks with both STFT-based and learnable signal representations, which correspond to frequency-domain and time-domain systems, respectively. For frequency-domain systems, we use 32~ms window size, 8~ms hop size, and Hanning window to generate the magnitude spectrograms as the input feature, and use 4 stacked Bi-LSTM layers with 256 hidden units in each of the LSTM layer to estimate the time-frequency masks of the two target speakers. For the time-domain systems, we use 2~ms window size, 1~ms hop size, and 64 kernels in the 1-D convolutional and transposed convolutional layers for signal encoder and decoder, respectively, and use 6 stacked DPRNN layers with Bi-LSTM layers for both intra-chunk and inter-chunk RNN modules. The number of hidden units in each of the LSTM layer is set to 128, and the chunk size in the DPRNN blocks is set to 100 frames. ReLU activation is used as the nonlinear activation function for mask estimation in both frequency-domain and time-domain systems.

For DPRNN models, the theoretical system latency equals to the chunk size, which we set to 100 frames (100 ms with 1 ms frame hop size). For RNN models, the theoretical system latency equals to the frame length (32 ms).

\subsection{Training and Evaluation}

For training, we use negative SNR between the separated waveforms and the target waveforms as the training objective. Permutation invariant training (PIT) is applied in all experiments \cite{yu2017permutation}. We use the Adam optimizer \cite{kingma2014adam} with the initial learning rate of 0.001, and we decay the learning rate by a factor of 0.5 if no best training model is found in three consecutive epochs. Gradient clipping by a maximum gradient norm of 5 is applied. We train the models until no best validation model is found in 15 consecutive epochs. For evaluation, the scale-invariant signal-to-distortion ratio improvement (SI-SDRi) \cite{le2019sdr} and signal-to-distortion ratio improvement (SDRi) \cite{vincent2006performance} are selected to measure the speech separation performance.

\section{Results and discussions}
\label{sec:result}
\subsection{Effect of Normalization Operations}

We first evaluate the effect of the modification of the normalization operations in offline models. This allows us to obtain a baseline system for fair comparison with online models trained with the same normalization operations. Table~\ref{tab:norm} shows the results of the offline and online models with different normalization operations, where ``Enc norm'' corresponds to the normalization operation applied to the signal encoder output, ``RNN norm'' corresponds to the normalization operation applied to the RNN layers for frequency-domain models and inter-chunk RNN modules for time-domain models, ``TD'' and ``FD'' stand for ``time-domain'' and ``frequency-domain'', respectively. ``RNN layers'' corresponds to the number of unidirectional LSTM layers used in each module, where ``1 layer'' corresponds to the decomposition scheme and ``2 layers'' corresponds to the reorganization scheme. For the 2-layer configuration, the outputs from the two LSTM layers are concatenated in a same way as the procedure in a standard Bi-LSTM layer. We notice that replacing the gLN operations in both encoder output and the RNN layers in the offline model by the cLN operations does not hurt the separation performance in TD models and even improves the performance in FD models, which indicates that the choice of a proper normalization operation may need further investigation. For comparison, the online systems use uni-directional LSTM layers in either the stacked RNN architecture or the inter-chunk RNN modules in the stacked DPRNN architecture, and up to 1.9 dB performance degradation compared to the offline system with same normalization operations is observed in terms of SI-SDRi. This confirms the observation in prior studies that when trained from scratch, online models can have a large performance gap compared to offline models with same basic model building blocks or architectures.

\begin{table}[!htbp]
    \centering
	\begin{tabular}{c|c|c|c|cc|cc}
		\thline
		Enc & RNN & RNN & \multirow{2}{*}{Online} & \multicolumn{2}{c|}{SI-SDRi} & \multicolumn{2}{c}{SDRi} \\
		norm & norm & layers & & TD & FD & TD & FD \\
		\hline
		gLN & gLN & -- & \texttimes & 9.5 & 9.1 & 10.3 & 10.0 \\
		cLN & cLN & -- & \texttimes & 9.5 & 9.4 & 10.3 & 10.3 \\
		\hline
		\multirow{2}{*}{cLN} & \multirow{2}{*}{cLN} & 1 & \multirow{2}{*}{\checkmark} & 7.6 & 7.6 & 8.8 & 8.8 \\
		 & & 2 & & 7.8 & 7.5 & 8.8 & 8.7 \\
		\thline
	\end{tabular}
	\caption{Separation performance of online and offline models with different normalization operations. All models are trained from scratch. SI-SDRi and SDRi are reported on decibel scale.}
	\label{tab:norm}
\end{table}

\subsection{Effect of Training Strategies}

We then evaluate the effect of initializing the online models with a pretrained offline model and training a same model to perform both online and offline separation. Table~\ref{tab:online} presents the results of the online models with different training strategies, where ``MT'' stands for the multitask training strategy where both online and offline outputs are used for training, and we report the performance of both online and offline paths for such models. ``D'' and ``R'' modes stand for ``decomposition'' and ``reorganization'' schemes, respectively. For online models that use the pretrained offline model for initialization, the performance of TD systems can be greatly improved compared to the baseline systems in Table~\ref{tab:norm} in both schemes, while the performance of the FD system only improves with the reorganization scheme. One possible explanation for this is that since the sequence length of the LSTM layers in FD systems is longer than that in TD systems and the DPRNN-based TD systems contain Bi-LSTM-based intra-chunk modules that are shared by both online and offline systems, the finetuning of TD systems is easier than that of FD systems. The performance of the offline path in models with multitask training are comparable to those in the baseline, which shows that both decomposition and reorganization schemes will not introduce severe performance degradation to the offline system. We also observe that models with the reorganization scheme have a consistently better online path performance in both TD and FD configurations, which indicates that the reorganization scheme might be a more general scheme in different architectures and model configurations. Moreover, TD models with stacked DPRNN modules achieve higher improvements than FD models with stack Bi-RNN layers, and the relative performance gap between the online and offline models is mitigated by up to 50\%.

\begin{table}[!htbp]
    \centering
	\begin{tabular}{c|c|c|c|cc|cc}
		\thline
		\multirow{2}{*}{Init} & \multirow{2}{*}{MT} & \multirow{2}{*}{Mode} & \multirow{2}{*}{Online} & \multicolumn{2}{c|}{SI-SDRi} & \multicolumn{2}{c}{SDRi} \\
		 & & & & TD & FD & TD & FD \\
		\hline
		\multirow{2}{*}{\checkmark} & \multirow{2}{*}{\texttimes} & D & \multirow{2}{*}{\checkmark} & 8.6 & 7.3 & 9.6 & 8.6 \\
		 & & R & & 8.8 & 8.0 & 9.7 & 9.2 \\
		 \hline
		\multirow{4}{*}{\texttimes} & \multirow{4}{*}{\checkmark} & \multirow{2}{*}{D} & \texttimes & 9.4 & 9.7 & 10.2 & 10.5 \\
		 & & & \checkmark & 8.3 & 8.0 & 9.4 & 9.2 \\
		 \cline{3-8}
		 & & \multirow{2}{*}{R} & \texttimes & 9.4 & 9.1 & 10.2 & 10.0 \\
		 & & & \checkmark & 8.3 & 8.3 & 9.4 & 9.4 \\
		 \hline
		 \multirow{4}{*}{\checkmark} & \multirow{4}{*}{\checkmark} & \multirow{2}{*}{D} & \texttimes & 9.8 & 9.4 & 10.6 & 10.2 \\
		 & & & \checkmark & 8.5 & 7.4 & 9.4 & 8.6 \\
		 \cline{3-8}
		 & & \multirow{2}{*}{R} & \texttimes & 9.8 & 9.5 & 10.6 & 10.4 \\
		 & & & \checkmark & 8.5 & 8.0 & 9.5 & 9.1 \\
		\thline
	\end{tabular}
	\caption{Separation performance of online models with different training strategies. SI-SDRi and SDRi are reported on decibel scale.}
	\label{tab:online}
\end{table}

\subsection{Discussions}

Prior works have been investigating the usage of knowledge distillation \cite{hinton2015distilling} to improve the performance of an online system by a large offline system \cite{aihara2019teacher, kurata2020knowledge, yu2020dual}. The way knowledge distillation is applied to the speech separation tasks can be broadly categorized to \textit{data augmentation} and \textit{output matching}, where the former one uses the large teacher model to generate pseudo training targets on a (typically) unlabelled dataset for the smaller student model \cite{zhang2021teacher, chen2021ultra}, and the latter one requires that either the intermediate or the final outputs of the student and teacher models should be as similar as possible \cite{aihara2019teacher, yu2020dual, chen2021ultra}. Although such knowledge distillation pipelines can be directly applied in our proposed layer decomposition and reorganization schemes, preliminary experiments show that using the output from the offline path as an auxiliary training objective for the online path results in worse performance than simply applying the multi-task training method. One possible reason for this observation might be that in order to let the student model match the output of the teacher model, how the model parameters should be shared between the student and teacher models and how the training objective should be designed need more investigation. We leave the application of knowledge distillation in our framework as a future work.

\section{Conclusion}
\label{sec:conclusion}
In this paper, we proposed a method for decomposing or reorganizing a Bi-RNN layer into an \textit{online path} and an \textit{offline path}, which allowed a Bi-RNN layer to perform both online and offline processing with a same set of model parameters. We further proposed two training strategies for the two schemes. Offline RNN-based neural separation systems can thus either serve as a pretrained model for initializing an online system, or be modified to jointly perform online and offline separation. Experiment results showed that the proposed layer decomposition and reorganization schemes can effectively mitigate the performance gap between two RNN-based offline separation models and their online variants.

\bibliographystyle{IEEEtran}
\bibliography{refs}

\end{document}